\documentclass{article}
\usepackage{ijcai24}

\usepackage{times}
\usepackage{soul}
\usepackage{url}
\usepackage[hidelinks]{hyperref}
\usepackage[utf8]{inputenc}
\usepackage[small]{caption}
\usepackage{graphicx}
\usepackage{amsmath}
\usepackage{amsthm}
\usepackage{booktabs}
\usepackage[ruled]{algorithm2e}
\usepackage[switch]{lineno}

\usepackage{multirow}
\usepackage{multicol}
\usepackage{threeparttable}
\usepackage{tabularx}
\usepackage{booktabs}
\usepackage{makecell}
\usepackage{colortbl}
\usepackage{graphicx}
\usepackage{amssymb}

\newcommand{\ie}{\textit{i.e., }}
\newcommand{\eg}{\textit{e.g., }}
\newcommand{\blue}[1]{\textcolor{black}{#1}}
\title{\blue{Safeguarding Medical Image Segmentation Datasets against \\ Unauthorized Training via Contour- and Texture-Aware Perturbations}}

\author{\rm{Xun Lin}$^{1}$\!, Yi Yu$^{2}$\thanks{Corresponding author.}\!, Song Xia$^{2}$\!, Jue Jiang$^{4}$\!, Haoran Wang$^{1}$\!, Zitong Yu$^{3}$\!, Yizhong Liu$^{1}$\!, \\Ying Fu$^{5}$\!, Shuai Wang$^{1}$\!, Wenzhong Tang$^{1}$\!, Alex Kot$^{2}$
\\
$^1$Beihang University ~\quad
$^2$Nanyang Technological University ~\quad
$^3$Great Bay University \\
$^4$Memorial Sloan Kettering Cancer Center ~\quad
$^5$Beijing Institute of Technology \\
{\tt\small \{linxun, wangshuai\}@buaa.edu.cn ~\quad yuyi0010@e.ntu.edu.sg}
}

\begin{document}

\maketitle

\begin{abstract}
\blue{
The widespread availability of publicly accessible medical images has significantly propelled advancements in various research and clinical fields. 
Nevertheless, concerns regarding unauthorized training of AI systems for commercial purposes and the duties of patient privacy protection have led numerous institutions to hesitate to share their images.
This is particularly true for medical image segmentation (MIS) datasets, where the processes of collection and fine-grained annotation are time-intensive and laborious.
Recently, Unlearnable Examples (UEs) methods have shown the potential to protect images by adding invisible shortcuts. These shortcuts can prevent unauthorized deep neural networks from generalizing.
However, existing UEs are designed for natural image classification and fail to protect MIS datasets imperceptibly as their protective perturbations are less learnable than important prior knowledge in MIS, \textit{e.g.,} contour and texture features. 
To this end, we propose an \textbf{U}nlearnable \textbf{Med}ical image generation method, termed \textbf{UMed}. UMed integrates the prior knowledge of MIS by injecting contour- and texture-aware perturbations to protect images. 
Given that our target is to only poison features critical to MIS, UMed requires only minimal perturbations within the ROI and its contour to achieve greater imperceptibility (average PSNR is 50.03) and protective performance (clean average DSC degrades from 82.18\% to 6.80\%).
}
\end{abstract}

\begin{figure}[t]
    \centering 
    \includegraphics[width=0.482\textwidth]{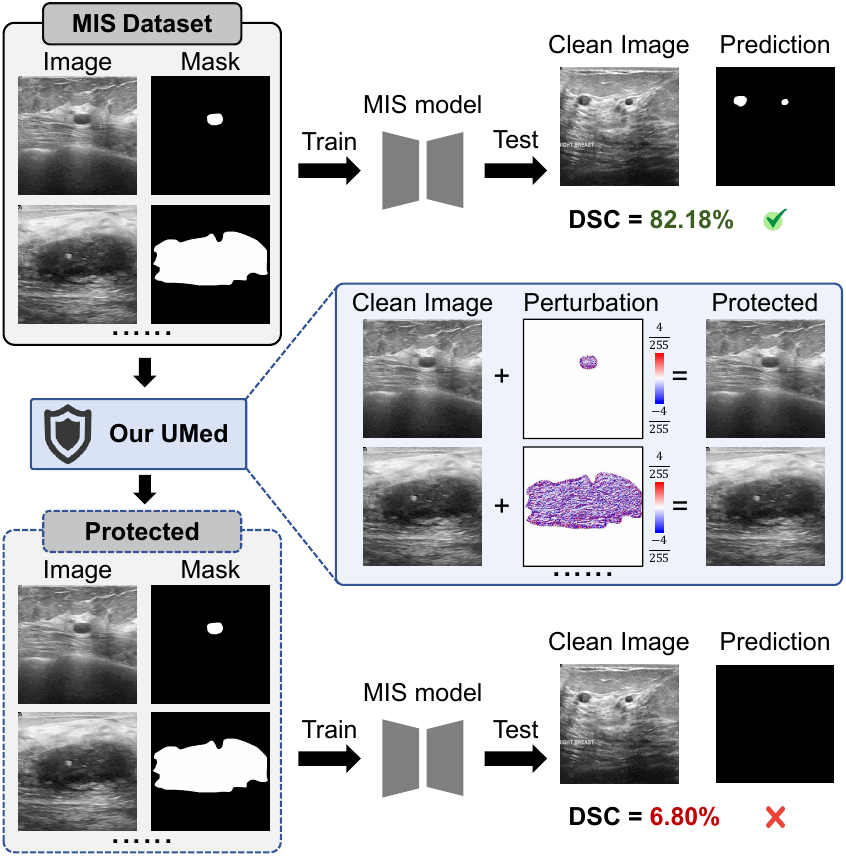}
        \caption{Illustration of using our proposed UMed to prevent an MIS dataset from unauthorized usage for AI model training. By adding protective perturbations to images of the MIS dataset, UMed can effectively reduce the clean segmentation performance of models trained on this dataset.}
    \label{fig:first}
\end{figure}

\section{Introduction}
\blue{
Medical images play a crucial role in the healthcare industry~\cite{dataset_importance}.
The proliferation of publicly available medical images has opened avenues for advancements in medical research and diagnosis~\cite{sharing}. 
However, this progress comes with the imperative to respect patient privacy and consent~\cite{consent,patient_privacy,privacy_2021_1}. 
Many patients prefer their medical images to be used solely for personal medical analysis and not for training AI models~\cite{personal,share_2012}, emphasizing the need for ethical handling of sensitive medical images. 
Furthermore, while some medical images are publicly shared for clinical educational purposes, such as guiding junior physicians in differentiating between tumors and normal tissues or familiarizing them with the anatomy of various organs \cite{education}, it is crucial to ensure that these images are not exploited for unauthorized commercial purposes \cite{federate_2022_1}. 
}

\blue{
Recently, Unlearnable Examples (UEs) techniques, which are special kinds of data poisoning attacks~\cite{biggio2012poisoning}, have been proposed for protecting images from unauthorized usage for AI model training.} 
By adding protective perturbations, which can be seen as easily learned shortcuts~\cite{lsp}, 
these techniques induce models trained on these safeguarded images to neglect the fundamental semantic content inherent within the images. Models trained on these protected images fail to make accurate predictions for clean images.
However, existing UEs methods are designed to protect natural image classification datasets, with an absence of exploration into safeguarding medical image segmentation (MIS) datasets.
For medical images, segmentation annotations are usually more time-consuming and need more expert knowledge than classification~\cite{seg_effort}. Unauthorized model training on collected MIS datasets may result in greater losses for the publishers. Due to the gap between the two tasks~\cite{task_gap}, previous UEs methods designed for classification, when directly applied to segmentation, lead to inadequate protection performance.
Meanwhile, given the high sensitivity of medical images to local details~\cite{adv_mia}, the invisibility of perturbations injected by existing methods is inadequate, and they pose a risk of altering the semantic content of the images~\cite{adv_attack}.

\begin{figure*}[t]
    \setlength{\abovecaptionskip}{4pt}
    \centering 
    \includegraphics[width=1.0\textwidth]{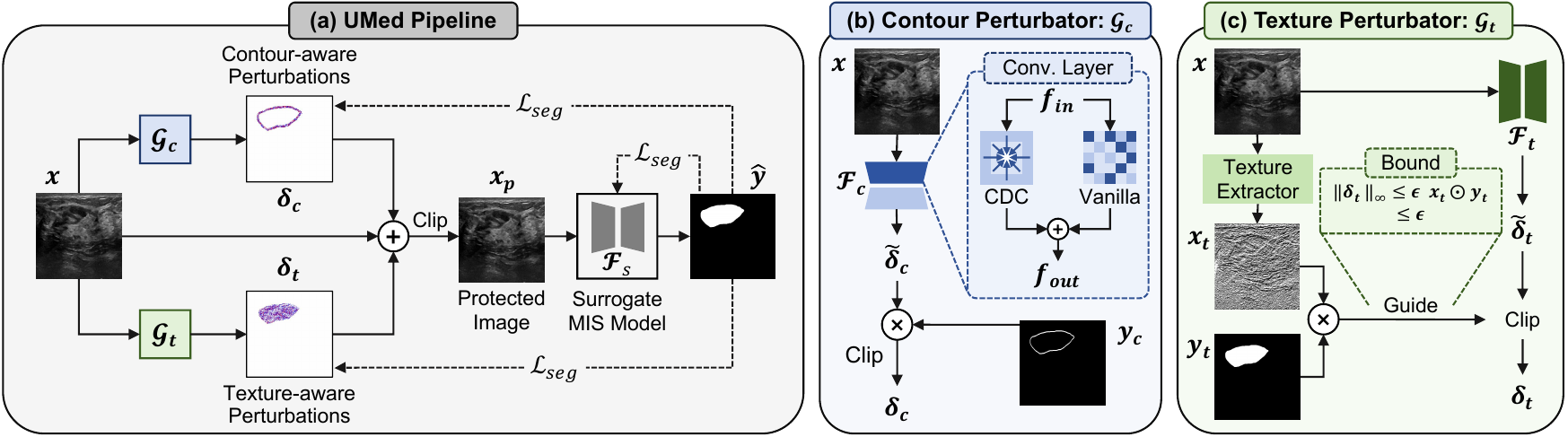}
    \caption{Illustration of the (a) pipeline of the proposed UMed, (b) contour perturbator $\mathcal G_c$ of UMed, and (c) texture perturbator $\mathcal G_t$ of UMed. $\mathcal G_c$ injects contour-aware perturbations using an encoder-decoder structured generator $\mathcal F_c$ integrated with central difference convolution (CDC) kernels. $\mathcal G_t$ perturbs textures within the ROI constrained by a texture-aware adaptive bound.} 
    \label{fig:umed}
\end{figure*}

Recent progress has discussed the importance of utilizing prior knowledge of contours and textures for MIS \cite{ife,boundary,xbound,freq_unet}. 
\blue{This inspires us to create more effective shortcuts by perturbing these two kinds of vital features.}
As presented in Fig.~\ref{fig:umed}(a), we propose a \textbf{U}Es generation method specifically for \textbf{Med}ical image segmentation (\textbf{UMed}), which focuses on poisoning the contours and internal textures of the region of interest (ROI). 
By considering these priors, our UMed can enhance the protection effectiveness while introducing fewer perturbations.

For the contour perturbator in UMed, we develop an encoder-decoder structured generator specifically for injecting perturbations into the contour regions. 
Given that the contours of the ROI often exhibit pixel intensity differences from their surrounding areas \cite{cdc_edge,cdc_fas}, we propose the central-difference-aware contour perturbator to encourage the capture and enhancement of these differences.
This contour perturbator can more effectively guide MIS models in learning the difference-enhanced contour perturbations while ignoring the original contour information. 
For the texture perturbator, different from existing methods \cite{em,tap,sep} which use a fixed bound to ensure perturbation invisibility, we employ a pixel-wise adaptive bound based on the texture feature intensities of each pixel for adding perturbations within the ROI, which can enhance perturbations' imperceptibility. Similar to the contour-aware perturbations, our texture perturbator creates a more easily learnable texture shortcut than the original texture features, leading MIS models to overlook the inherent texture characteristics during training.

Our contributions are summarized as follows.
\begin{itemize}
    \item To the best of our knowledge, we propose the first unlearnable examples generation method, namely UMed, specifically designed for MIS datasets.
    \item \blue{With the consideration of high imperceptibility requirements of medical images, we find that adding smartly designed contour- and texture-aware perturbations within the ROI not only reduces visibility but also more effectively protects MIS datasets.}
    \item \blue{Our strategies of generating protective perturbations based on texture and contour priors offer a novel and perturbation-efficient way of creating unlearnable examples for future research.}
\end{itemize}

\section{Related Works}
\subsection{Medical Image Segmentation}
MIS is an important step in medical diagnosis and treatments, aiming to locate ROIs such as tumors, lesions, anatomical structures, or organs at the pixel-level \cite{mis_review}.
In the early years, with the boost of deep Convolutional Neural Networks (CNNs), FCN successfully introduces the CNN for image segmentation~\cite{fcn}.
Among CNN-based methods, U-Net \cite{unet} has gained significant attention due to the effectiveness of the proposed U-shaped encoder-decoder structure and skip connections. Motivated by U-Net, a set of U-Net-based methods are developed \cite{unet_review,unet_review_access}.
U-Net++ \cite{unetplus} proposes nested and dense skip connections and increases the depth or width of the backbone network for better representation learning.
Subsequently, motivated by the success of attention mechanisms and transformers, Attention U-Net \cite{attnunet} and TransUNet \cite{transunet} integrate these techniques for better capturing global correlations within medical images.
More recently, some useful prior knowledge of MIS, \eg contour \cite{boundary,ambigous,xbound,connective} and texture features \cite{ife,implicit,freq_unet} are discussed and employed to further enhance their backbone networks.

\subsection{Unlearnable Examples}
Unlearnable examples (UEs), as a specific type of data poisoning attacks~\cite{barreno2010security,goldblum2022dataset}, aim to protect the datasets from unauthorized model training by
applying subtle modifications (\textit{e.g.,} bounded perturbations $\Vert \pmb{\delta} \Vert_{\infty} \le \frac{8}{255}$) to images from the entire training dataset with correct labels. 
UEs are considered a promising approach for data protection, leading models trained on such datasets to approach random guessing performance on clean test data.
Noteworthy techniques include EM~\cite{em}, which employs error-minimizing noise as perturbations, {AR~\cite{ar} using an autoregressive poisoning method with a manually-specified CNN}, and TAP~\cite{tap} employing targeted adversarial examples as unlearnable examples. LSP~\cite{lsp} explores efficient and surrogate-free unlearnable examples, extending perturbations to be $\ell_2$ bounded. {SEP~\cite{sep} ensemble the checkpoints of the surrogate model as diverse data protectors, enhancing transferability and protectiveness.
However, these UEs methods, designed to protect image classification datasets, cannot effectively protect MIS datasets and fail to meet the invisibility requirements of perturbations in medical images.}

\section{Method}
\subsection{Preliminary}
\noindent\textbf{Medical Image Segmentation.}
Given a clean dataset $\mathcal D_{clean} = \{(\boldsymbol{x}_i, \boldsymbol{y}_i)\}_{i=1}^n$ for training, where $\boldsymbol{x}_i\in \mathbb{R}^{H\times W \times C}$ is the input medical image and $\boldsymbol{y}_i\in \{0, 1\}^{H\times W}$ is the corresponding binary map of the ROI in $\boldsymbol{x}_i$. MIS aims to optimize a neural network $\mathcal F(\cdot;\theta)$ to build the mapping relationship from $\boldsymbol{x}$ to $\boldsymbol{y}$. This optimization can be formulated as follows
\begin{equation}
    \underset{\theta}{\text{argmin }}\mathbb{E}_{(\boldsymbol{x}, \boldsymbol{y})\sim D_{clean}}\Big[\mathcal L_{seg}(\mathcal F(\boldsymbol{x}; \theta), \boldsymbol{y})\Big],
\end{equation}
where $\mathcal L_{seg}$ is the loss function (\eg cross-entropy loss and dice loss) and $\theta$ is the trainable parameters of $\mathcal F$.

\vspace{1.5mm}
\noindent\textbf{Unlearnable Examples.}
UEs are generated to prevent unauthorized MIS models $\mathcal F$ from learning any useful information by adding invisible perturbations to input images in training datasets $\mathcal D_{clean}$. For better understanding, we take the best-know UEs method, \textit{i.e.,} error-minimizing (EM) noise, proposed in \cite{em} as an example. EM adopts a bi-objective optimization to generate protective perturbations, which are given by
\begin{equation}
    \underset{\theta}{\text{min }}\mathbb{E}_{(\boldsymbol{x}, \boldsymbol{y})\sim D_{clean}}\Big[\underset{\boldsymbol{\delta} \in \mathcal{I}}{\text{min }}\mathcal L_{seg}(\mathcal F_s(\boldsymbol{x}+\boldsymbol{\delta}; \theta), \boldsymbol{y})\Big],
    \label{em_generation}
\end{equation}
where $\mathcal F_s$ is a surrogate model, $\mathcal{I}$ is the feasible regions for the perturbations (\eg $\left \| \boldsymbol\delta \right \|_{\infty} \leq \epsilon$ for invisibility), and $\boldsymbol{x}+\boldsymbol{\delta}$ is the protected images and it is also called the unlearnable examples. In this process, $\theta$ and $\boldsymbol{\delta}$ are alternately optimized to minimize $\mathcal L_{seg}$.
Finally, an unlearnable dataset $\mathcal D_{protect}=\{(\boldsymbol{x}_i+\boldsymbol{\delta}_i, \boldsymbol{y}_i)\}_{i=1}^n$ is derived.

\vspace{1.5mm}
\noindent\textbf{Data Protector and Data Exploiter.}
In this work, we act as a data protector, aiming to safeguard the dataset $\mathcal D_{clean}$, and generate a protected dataset $\mathcal D_{protect}$ from which learning useful information for segmentation is difficult. 
The data exploiter's goal is to train their MIS model $\mathcal F$ based on this $\mathcal D_{protect}$ and learn as much useful information as possible, achieving good performance on other clean images. 
We consider a scenario close to reality to constrain the capabilities of the protector. 
These constraints are: (1) the protector typically does not know which segmentation model the exploiter will use for training, we restrict the protector to only use one surrogate model for generating perturbations, and (2) since protectors usually have the opportunity to manipulate images before dataset release, we allow the protector to have complete control over $\mathcal D_{clean}$.

\subsection{Our UMed}
In this section, we present the pipeline of using the proposed UMed to generate unlearnable MIS datasets. 
Then, we describe how UMed perturbs crucial features for training MIS models by generating the contour- and texture-aware perturbations, respectively.
Finally, we introduce the detailed optimization strategy of UMed.

\vspace{1.5mm}
\noindent\textbf{Overview.}
UMed generates perturbations based on contour and texture features that are crucial for MIS, aiming to enhance protection performance while reducing the visibility of perturbations. 
As shown in Fig.~\ref{fig:umed}(a), UMed consists of two branches utilizing the proposed contour perturbator $\mathcal G_c$ and texture perturbator $\mathcal G_t$ to generate contour- and texture-aware perturbations $\boldsymbol{\delta}_c$ and $\boldsymbol{\delta}_t$, respectively. 
{
Existing sample-wise UEs methods, which optimize perturbations directly for each image~\cite{em,tap,sep}, fail to take into account the distribution characteristics of the MIS dataset. This makes them less effective and generalizable in perturbing features, thus resulting in the lack of transferability against different MIS models. To address this issue, we propose the trainable encoder-decoder structured generators $\mathcal F_c$ and $\mathcal F_t$ for learning more robust protective perturbations from MIS datasets.
}
Upon optimizing UMed, $\boldsymbol{\delta}_c$ and $\boldsymbol{\delta}_t$ are produced and added to the original image $\boldsymbol{x}$ for protection. To ensure the imperceptibility of the perturbations, following previous works \cite{em,tap}, we constrain them with a $\ell_{\infty}$-norm bound, setting $\epsilon$ to be $\frac{4}{255}$. 

\vspace{1.5mm}
\noindent\textbf{Contour Perturbator.}
\blue{
In medical images, the contour of the ROI always displays differences in pixel intensity relative to the neighboring pixels of the contour. Such differences, whether subtle or pronounced, serve as crucial cues for MIS models' predictions.
To create shortcuts on contours, we aim to generate perturbations that enhance the differences between contours and their surrounding pixels. This strategy ensures that these enhanced contours are easier to learn than the original ones.
Inspired by the special convolution operations designed for contour-sensitive vision tasks \cite{cdc_edge,cdc_fas}, which capture the differences between the central pixel of a kernel and its surroundings to enhance contour representation, we propose the contour perturbator $\mathcal G_c$ presented in Fig.~\ref{fig:umed}(b). 
Specifically, we improve the pixel difference perceptibility of generator $\mathcal F_c$ by integrating central-difference-aware kernels into all convolutional layers of its encoder. Taking $f_{in}$ and $f_{out}$ as the input and output feature maps, respectively, these convolution layers can be formulated as follows
\begin{equation}
    \begin{aligned}
       f_{out}(r) & = \underbrace{\sum_{\Delta r \in \mathcal R} w_v(\Delta r)\cdot f_{in}(r-\Delta r)}_{\text{Vanilla convolution}} \\
        & + \underbrace{\sum_{\Delta r \in \mathcal R} w_c(\Delta r)\cdot\big[f_{in}(r-\Delta r) - f_{in}(r)\big]}_{\text{Central difference convolution}},
    \end{aligned}
\end{equation}
where $r$ denotes the current position of the convolution kernel conducting on $f_{in}$ and $f_{out}$, and $\mathcal R=\{(1,\!1),(0,\!1),\cdots, (-1,\!0), (-1,\!-1)\}$ is the local respective field of the kernel. $w_v$ and $w_c$ are the trainable $3\!\times\!3$ kernels for the two types of convolution operations, respectively.
Subsequently, the process of generating contour-aware perturbations can be described as follows
\begin{equation}
    \boldsymbol{\delta}_c = \mathcal G_c(\boldsymbol{x}) = \text{Clip}_{[-\epsilon, \epsilon]}\big[\mathcal F_c(\boldsymbol{x})\odot \boldsymbol{y}_c\big],
\end{equation}
where $\boldsymbol{y}_c \in \{0, 1\}^{H\times W}$ represents the binary map of contour regions in $\boldsymbol{x}$, and a clipping operation \textbf{Clip} is employed to guarantee that the perturbations are subject to $\left \| \boldsymbol{\delta}_c \right \|_{\infty} \leq \epsilon$.
}

\vspace{1.5mm}
\noindent\textbf{Texture Perturbator.}
Texture is also a crucial prior knowledge for MIS, where the inconsistency between the textures inside and outside of the ROI often provides useful cues for many MIS methods \cite{ife,lbp_mia}. 
Our objective here is to introduce subtle perturbations within the ROI when preserving the natural appearance of the texture inside the ROI. 
Simultaneously, the added texture-aware perturbations are designed to be more easily learnable for models used by data exploiters compared to the original texture features of the ROI. 
As the diverse grayscale ranges present in medical images acquired through various imaging techniques, we utilize the grayscale-independent Local Binary Patterns (LBP) descriptor \cite{lbp_mia} as our texture feature extractor.
As illustrated in Fig. \ref{fig:umed}(c), we first generate initial perturbations $\boldsymbol{\tilde{\delta}}_t$  for $\boldsymbol{x}$ using the generator $\mathcal F_t$ with an encoder-decoder structure. Meanwhile, we extract the texture intensity for each pixel using the texture feature extractor. For further invisibility of the perturbations, we utilize only the texture features within the ROI and add perturbations solely within it. After obtaining the texture feature map $\boldsymbol{x}_{t} \in \mathbb R^{H\times W}$ inside the ROI, we guide $\mathcal F_t(\boldsymbol{x})$ to perform adaptive clipping, \textit{i.e.} $\|\boldsymbol{\delta}_t\|_{\infty} \leq \epsilon \cdot \boldsymbol{x}_{t}\odot \boldsymbol{y}_t$, allocating higher perturbation bound to areas with strong texture features while reducing it in regions with weak texture features. 
Here $\boldsymbol{y}_t = (1 - \boldsymbol{y}_c) \odot \boldsymbol{y} \in \{0, 1\}^{H\times W}$ denotes the binary map of regions within the ROIs, excluding the contour. Consequently, the perturbations are confined to the regions solely within the ROIs.
And the final perturbations are given by
\begin{equation}
    \boldsymbol{{\delta}}_t = \mathcal G_t(\boldsymbol{x}) = \text{Clip}_{[-\epsilon \cdot \boldsymbol{x}_{t}\odot \boldsymbol{y}_t, \epsilon \cdot \boldsymbol{x}_{t}\odot \boldsymbol{y}_t]}\big[\mathcal F_t(\boldsymbol{x})\big].
\end{equation}
It is important to note that the perturbation bound for each pixel remains below the preset $\epsilon$. This approach allows us to create more easily learnable and less perceptible texture replicas within the ROIs under a reduced bound.

\begin{algorithm}[t]
\caption{Optimizing Process of UMed}\label{algorithm}
\SetKwInOut{Input}{Input}\SetKwInOut{Output}{Output}
\Input{Surrogate model $\mathcal F_s(\cdot;\theta_{\mathcal F_s})$, contour perturbator $\mathcal G_c(\cdot;\theta_{\mathcal G_c})$, texture perturbator $\mathcal G_t(\cdot;\theta_{\mathcal G_t})$, learning rates $\alpha_s$ and $\alpha_g$, number of epochs $e$, clean dataset $\mathcal D_{clean} = \{(\boldsymbol{x}_i, \boldsymbol{y}_i)\}_{i=1}^n$}
\vspace{1mm}
\Output{Optimized perturbators $\mathcal G_c(\cdot;\theta_{\mathcal G_c})$, $\mathcal G_t(\cdot;\theta_{\mathcal G_t})$}
\vspace{1mm}
\SetKwData{ModelParam}{\small model.parameters}
\SetKwFunction{Clip}{\small { Clip}}
\SetKwFunction{GradMod}{\small {ReGrad}}
\SetKwFunction{trainable}{\small {IsTrainable}}

\For{$i \leftarrow 1$ \textbf{to} $e$ }{
    \For{$\{\boldsymbol{x},\boldsymbol{y}\} \in \mathcal D_{clean}$}{
        $\boldsymbol{\delta}_c$, $\boldsymbol{\delta}_t \leftarrow \mathcal G_c(\boldsymbol{x})$, $\mathcal G_t(\boldsymbol{x})$\;
        $\boldsymbol{x}_p \leftarrow$$\text{Clip}_{[0,1]}\big[\boldsymbol{x} + \boldsymbol{\delta}_c + \boldsymbol{\delta}_t\big]$\;
        \If{$i$ \% 5 $\neq0$}{
            $\theta_{\mathcal G_c}\leftarrow \theta_{\mathcal G_c}-\alpha_g\cdot\nabla_{\theta_{\mathcal G_c}}\mathcal L_{seg}(\mathcal F_s(\boldsymbol{x}_p), \boldsymbol{y})$\;
            $\theta_{\mathcal G_t}\leftarrow \theta_{\mathcal G_t}-\alpha_g\cdot\nabla_{\theta_{\mathcal G_t}}\mathcal L_{seg}(\mathcal F_s(\boldsymbol{x}_p), \boldsymbol{y})$\;
        }
        \Else{
            $\theta_{\mathcal F_s}\leftarrow \theta_{\mathcal F_s}-\alpha_s\cdot\nabla_{\theta_{\mathcal F_s}}\mathcal L_{seg}(\mathcal F_s(\boldsymbol{x}_p), \boldsymbol{y})$\;
        }
    }
}
\label{alg:train}
\end{algorithm}

\begin{table*}[t]
    \setlength{\tabcolsep}{3.5pt}
    \centering
    \resizebox*{1.0 \linewidth}{!}{
        \begin{tabular}{@{}cccccccccc@{}}
        \toprule
        \multirow{2}{*}{\textbf{Protector}} & \multicolumn{1}{c}{\multirow{2}{*}{\textbf{Venue}}} & \multicolumn{2}{c}{\textbf{BUSI}} & \multicolumn{2}{c}{\textbf{Chest X-ray}} & \multicolumn{2}{c}{\textbf{Kvasir-SEG}} & \multicolumn{2}{c}{\textbf{Average}} \\ 
        \cmidrule(lr){3-4} \cmidrule(lr){5-6} \cmidrule(lr){7-8} \cmidrule(lr){9-10}
         & \multicolumn{1}{l}{} & \textbf{Jac.(\%) $\downarrow$} & \textbf{DSC(\%) $\downarrow$} & \textbf{Jac.(\%) $\downarrow$} & \textbf{DSC(\%) $\downarrow$} & \textbf{Jac.(\%) $\downarrow$} & \textbf{DSC(\%) $\downarrow$} & \textbf{Jac.(\%) $\downarrow$} & \textbf{DSC(\%) $\downarrow$} \\ 
         \midrule
         
        None & - & 58.15 & 72.85 & 84.93 & 91.85 & 69.48 & 81.83 & 70.85 & 82.18 \\
        EM \cite{em} & ICLR & 19.88 & 32.55 & 47.76 & 64.62 & 17.73 & 30.06 & 28.46 & 42.41 \\
        TAP \cite{tap} & NeurIPS & 21.56 & 34.77 & 85.73 & 92.32 & 14.94 & 25.84 & 40.74 & 50.98 \\
        LSP \cite{lsp} & SIGKDD & 20.72 & 34.16 & 81.58 & 89.85 & 48.06 & 64.38 & 50.12 & 62.80 \\
        AR \cite{ar} & NeurIPS & 25.81 & 40.39 & 84.26 & 91.45 & 22.71 & 36.75 & 44.26 & 56.20 \\
        SEP \cite{sep} & ICLR & 22.05 & 35.47 & 82.31 & 90.29 & 46.82 & 63.45 & 50.39 & 63.07 \\
        \textbf{UMed} & \textbf{Ours} & \textbf{0.46} & \textbf{0.92} & \textbf{10.76} & \textbf{19.24} & \textbf{0.12} & \textbf{0.23} & \textbf{3.78} & \textbf{6.80} \\ \bottomrule
        \end{tabular}
    }
    \caption{\blue{Comparison of protection performance (measured on clean testing images) among different protectors. Experiments are conducted with U-Net as the MIS model on three MIS datasets, \ie BUSI, Chest X-ray, and Kvasir-SEG.}}
    \label{tab:comp}
\end{table*}

\vspace{1.5mm}
\noindent\textbf{Optimizing UMed.}
The objective of UMed is to optimize the contour- and texture-aware perturbators $\mathcal G_c$ and $\mathcal G_t$, such that $\mathcal G_c(\boldsymbol{x}) +\mathcal G_t(\boldsymbol{x})+\boldsymbol{x}$ is more easily to learn by the data exploiters' models compared to $\boldsymbol{x}$ alone. During optimizing, we employ the loss $\mathcal L_{seg}$ with an equally weighted combination of cross-entropy loss and dice loss to optimize the surrogate MIS model $\mathcal F_s$ and our perturbators. 
\begin{equation}
\begin{split}
    & \boldsymbol{x}_p = \text{Clip}_{[0, 1]}\big[\mathcal G_c(\boldsymbol{x})+\mathcal G_t(\boldsymbol{x})+\boldsymbol{x}\big],\\
    & \mathcal L_{seg} = \mathcal L_{ce}(\mathcal F_s(\boldsymbol{x}_p), \boldsymbol{y}) + \mathcal L_{dice}(\mathcal F_s(\boldsymbol{x}_p), \boldsymbol{y}).
\end{split}
\label{seg_loss}
\end{equation}
Subsequently, the optimization process is given by
\begin{equation}
    \underset{\theta_{\mathcal F_s}}{\text{min }}\mathbb{E}_{(\boldsymbol{x}, \boldsymbol{y})\sim D_{clean}}\Big[\underset{\theta_{\mathcal G_c}, \theta_{\mathcal G_t}}{\text{min }}\mathcal L_{seg}\Big],
\end{equation}
where $\mathcal{L}_{seg}$ in defined in Eq.~(\ref{seg_loss}). 
In contrast to EM~\cite{em} as defined in Eq.~(\ref{em_generation}), or other UEs such as TAP~\cite{tap} and SEP~\cite{sep}, which generate perturbations across entire regions with identical bounds for each pixel, our approach incorporates prior information encompassing both contour and texture. This effectively restricts the search space for perturbations, and results in more imperceptibly protected images. Subsequent experiments demonstrate that with this more constrained perturbation space, our proposed method achieves superior performance in preventing unauthorized model training on the datasets.
In Algorithm \ref{alg:train}, we present the detailed process of alternately training $\mathcal F_s$ and the two perturbators $\mathcal G_c$ and $\mathcal G_t$. We employ the widely used medical image segmentation model, \ie U-Net \cite{unet}, as the surrogate model and also as the perturbation generator $\mathcal F_t$ within $\mathcal G_t$.

\section{Experimental Results}
\subsection{Experimental Setups}
\noindent\textbf{Datasets.}
To comprehensively validate the effectiveness of UMed, we select three widely used image segmentation datasets, each acquired with different imaging devices and capturing distinct subjects. These datasets are BUSI (breast ultrasound tumor segmentation, 612 images) \cite{BUSI}, Kvasir-SEG (endoscopic polyp segmentation, 1000 images) \cite{kvasir}, and Chest X-ray (lung segmentation, 704 images) \cite{lung_dataset1,lung_dataset2}. Note that Kvasir-SEG is a three-channel dataset, whereas BUSI and Chest X-ray are single-channel. We ensure that the number of channels in the added perturbations matches the channel count of the input images. For the following experiments, we randomly split the datasets into training and testing sets, maintaining a ratio of 8:2.

\begin{table*}[t]
\setlength{\tabcolsep}{11.9pt}
\centering
\renewcommand{\arraystretch}{0.9}
\scalebox{0.78}{
    \begin{tabular}{@{}c|ccccccccc@{}}
    \toprule
    \multirow{2}{*}{\textbf{Dataset}} & \textbf{Exploiter} $\ \rightarrow$ & \multicolumn{2}{c}{\textbf{Attention U-Net}} & \multicolumn{2}{c}{\textbf{U-Net++}} & \multicolumn{2}{c}{\textbf{TransUNet}} & \multicolumn{2}{c}{\textbf{Average}} \\ 
    \cmidrule(lr){2-2} \cmidrule(lr){3-4} \cmidrule(lr){5-6} \cmidrule(lr){7-8} \cmidrule(lr){9-10}
     & \textbf{Protector} $\ \downarrow$  & \textbf{Jac.(\%)$\ \downarrow$} & \textbf{DSC(\%)$\ \downarrow$} & \textbf{Jac.(\%)$\ \downarrow$} & \textbf{DSC(\%)$\ \downarrow$} & \textbf{Jac.(\%)$\ \downarrow$} & \textbf{DSC(\%)$\ \downarrow$} & \textbf{Jac.(\%)$\ \downarrow$} & \textbf{DSC(\%)$\ \downarrow$} \\ \midrule
    \multirow{7}{*}{\rotatebox{90}{\textbf{BUSI}}} & None & 56.95 & 72.22 & 59.02 & 73.81 & 56.68 & 72.06 & 57.55 & 72.70 \\
     & EM & 18.50 & 31.00 & 54.94 & 70.52 & 54.00 & 69.70 & 42.48 & 57.07 \\
     & TAP & 11.10 & 19.73 & 54.79 & 70.06 & 51.34 & 67.41 & 39.08 & 52.40 \\
     & LSP & 10.36 & 18.67 & 58.70 & 73.37 & 55.42 & 70.82 & 41.49 & 54.29 \\
     & AR & 30.77 & 46.91 & 58.86 & 73.75 & 56.35 & 71.81 & 48.66 & 64.16 \\
     & SEP & 24.41 & 39.00 & 39.11 & 54.93 & 49.85 & 66.20 & 37.79 & 53.38 \\
     & \textbf{UMed} & \textbf{0.68} & \textbf{1.35} & \textbf{4.31} & \textbf{7.84} & \textbf{8.33} & \textbf{15.26} & \textbf{4.44} & \textbf{8.15} \\ \midrule
    \multirow{7}{*}{\rotatebox{90}{\textbf{Chest X-ray}}} & None & 88.90 & 94.12 & 91.43 & 95.52 & 92.52 & 96.11 & 92.21 & 95.95 \\
    & EM & 53.81 & 69.95 & 90.50 & 95.01 & 91.07 & 95.32 & 78.46 & 86.76 \\
     & TAP & 85.73 & 92.32 & 91.36 & 95.48 & 92.17 & 95.92 & 89.75 & 94.57 \\
     & LSP & 82.06 & 90.14 & 91.26 & 95.43 & 92.43 & 96.07 & 88.58 & 93.88 \\
     & AR & 78.26 & 87.80 & 91.03 & 95.30 & 92.48 & 96.09 & 87.26 & 93.06 \\
     & SEP & 92.28 & 95.98 & 91.42 & 95.51 & 90.66 & 95.61 & 91.45 & 95.70 \\
     & \textbf{UMed} & \textbf{13.30} & \textbf{23.44} & \textbf{19.78} & \textbf{31.89} & \textbf{43.98} & \textbf{60.81} & \textbf{25.69} & \textbf{38.71} \\ 
     \midrule
     \multirow{7}{*}{\rotatebox{90}{\textbf{Kvasir-SEG}}} & None & 69.83 & 81.92 & 67.00 & 80.11 & 69.12 & 81.61 & 68.65 & 81.21 \\
     & EM & 33.43 & 49.22 & 59.06 & 73.87 & 64.29 & 77.91 & 52.26 & 67.00 \\
     & TAP & 49.52 & 62.50 & 58.08 & 73.07 & 58.62 & 73.74 & 55.41 & 69.77 \\
     & LSP & 59.99 & 74.66 & 62.01 & 76.18 & 64.96 & 78.70 & 62.32 & 76.51 \\
     & AR & 65.42 & 78.97 & 61.12 & 75.67 & 65.47 & 78.95 & 64.00 & 77.86 \\
     & SEP & 47.29 & 63.72 & 56.65 & 71.37 & 55.43 & 70.98 & 53.12 & 68.69 \\
     & \textbf{UMed} & \textbf{0.67} & \textbf{1.31} & \textbf{2.76} & \textbf{5.22} & \textbf{11.23} & \textbf{19.57} & \textbf{4.89} & \textbf{8.70} \\ \bottomrule
    \end{tabular}
}
\caption{Comparison of transferability (measured on clean testing images) among different protectors. The protector uses U-Net as the surrogate model, whereas the exploiter adopts three different MIS models, \ie Attention U-Net, U-Net++, and TransUNet.}
\label{tab:trans}
\end{table*}

\begin{figure*}[t]
    \centering 
    \includegraphics[width=1.0\textwidth]{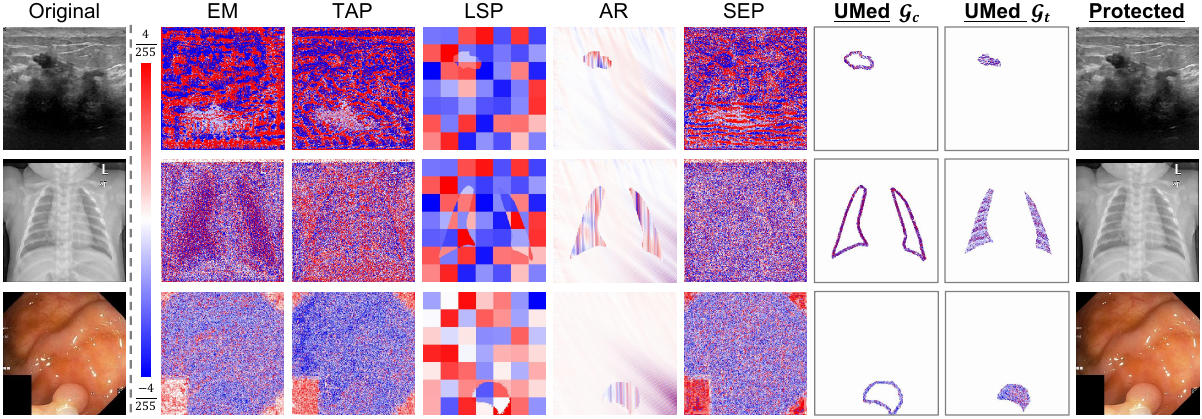}
    \caption{{Visualization of the perturbations generated by different protectors. From left to right, each column represents original images, perturbations generated by EM, TAP, LSP, AR, SEP, $\mathcal G_c$ of UMed, $\mathcal G_t$ of UMed, and the images protected by UMed, respectively.}}
    \label{fig:vis}
\end{figure*}

\vspace{1.5mm}
\noindent\textbf{Competing UE Methods (Protectors).}
We compare our UMed with two types of UEs methods. The first type requires a surrogate model to optimize perturbations, \ie EM \cite{em}, TAP \cite{tap}, and SEP~\cite{sep}. The second type is model-agnostic and class-wise, \ie LSP \cite{lsp} and AR \cite{ar}. To adapt LSP and AR for MIS tasks, we treat pixels inside and outside the ROI as two distinct classes.

\vspace{1.5mm}
\noindent\textbf{Training of MIS Models (Exploiters).}
We utilized four popular MIS models, namely U-Net \cite{unet}, U-Net++ \cite{unetplus}, Attention U-Net \cite{attnunet}, and TransUNet \cite{transunet}, to simulate the models that data exploiters might employ for training on the protected datasets. The number of training epochs is set to 150 with an Adam optimizer (using fixed lr=1e-4) and a batch size of 32. The loss function we utilize here is also $\mathcal L_{seg}$.

\vspace{1.5mm}
\noindent\textbf{Evaluation Metrics.}
We evaluate the protective capability of the protectors using MIS metrics, \ie the Dice Similarity Coefficient (DSC) and Jaccard similarity (Jac.), computed on clean testing datasets. To measure the invisibility of the perturbations, we calculate the average Peak Signal-to-Noise Ratio (PSNR) and the Structural Similarity Index (SSIM) between every protected image and its original counterpart.

\vspace{1.5mm}
\noindent\textbf{Implementation Details of UMed.}
In real-world scenarios where the specific model a data exploiter might employ for training is unknown, we default to using U-Net as the surrogate model to optimize the perturbators.
Given the sensitivity of medical images to perturbations \cite{adv_mia,adv_attack}, which is distinct from methods based on natural images where the perturbation bound in terms of $\ell_\infty$-norm is commonly set to $\frac{8}{255}$, we impose a more stringent bound of $\frac{4}{255}$ for our method. The size of the images is standardized to $224\!\times\!224$. We use the Adam optimizer with learning rates $\alpha_s$ and $\alpha_g$, both set to $10^{-4}$. The batch size is 16. The number of epochs for the surrogate model training is set to 100. After each training epoch of the surrogate model, the two generators are trained for four epochs each.

\subsection{Analysis on the Protection Capability}
\noindent\textbf{Effectiveness.} 
We conduct experiments with both the surrogate model and the data exploiters' MIS models to be U-Net. We train U-Net on the protected training dataset and test them on clean testing datasets. Results in Table \ref{tab:comp} demonstrate that our UMed has significantly superior protection capabilities across different datasets, with an average Jac. of 3.78\% and DSC of 6.80\%, greatly outperforming existing algorithms such as the best-known EM, which has an average Jac. of 28.46\% and DSC of 42.41\%. 
Remarkably, UMed achieves inspiring protection on BUSI and Kvasir-SEG, with both segmentation metrics on clean testing datasets nearly dropping to zero. 
Additionally, while other protectors show reasonable performance on BUSI and Kvasir-SEG, they fail to protect Chest X-ray well.
This is due to the ROIs in the Chest X-ray dataset are the left and right lungs, which have a strong positional prior.
Such positional priors are easily learned by models. Other protectors, which ignore the priors of MIS during perturbation generation, are less effective—even though the perturbations generated by methods like EM, TAP, and SEP are optimized. 
In contrast, our UMed focuses on poisoning the contours and textures of the ROI, creating shortcuts that are more easily learnable than these strong positional priors, thus offering more protective capability.

\vspace{1.3mm}
\noindent\textbf{Transferability.}
Since data exploiters' models are usually unknown, it is crucial to validate the transferability of these UEs generators. We further conduct experiments when the surrogate model is different from data exploiters' models. As shown in Table \ref{tab:trans}, 
the comparative methods have limited transferability. When data exploiters utilize models different from the surrogate model on each dataset, these protectors' clean Jac. and DSC significantly drop. In contrast, our UMed, by introducing prior knowledge that can generalize well and is easily learnable by various models, achieves the best transferability.
Furthermore, when the data exploiter adopts TransUNet, other protectors almost cannot offer protection. UMed's protective capability also drops on the Chest X-ray. This is due to the integration of a transformer within TransUNet, which incorporates position embeddings and primarily focuses on extracting global relationships. Given that positional priors are exceptionally easy to learn on this dataset, the perturbations generated by protectors for Chest X-ray are more difficult to learn by TransUNet compared to the inherent positional information. Despite this, our method still outperforms existing protectors. On other datasets without such strong priors, the protection performance of UMed is consistent among different models.

\begin{table*}[t]
   \centering
    \setlength\tabcolsep{7.4pt}
\scalebox{0.78}{
\begin{tabular}{@{}cccccccccccc@{}}
    \toprule
    \multirow{2}{*}{\textbf{Protector}} & \multirow{2}{*}{\textbf{Bound}} & \multicolumn{2}{c}{\textbf{w/o Defense}} & \multicolumn{2}{c}{\textbf{Gaussian Blur}} & \multicolumn{2}{c}{\textbf{JPEG Compression}} & \multicolumn{2}{c}{\textbf{Adv. Training}} & \multicolumn{2}{c}{\textbf{Invisibility}} \\ \cmidrule(lr){3-4} \cmidrule(lr){5-6} \cmidrule(lr){7-8} \cmidrule(lr){9-10} \cmidrule(l){11-12} 
     &  & \textbf{Jac.(\%) $\downarrow$} & \textbf{DSC(\%) $\downarrow$} & \textbf{Jac.(\%) $\downarrow$} & \textbf{DSC(\%) $\downarrow$} & \textbf{Jac.(\%) $\downarrow$} & \textbf{DSC(\%) $\downarrow$} & \textbf{Jac.(\%) $\downarrow$} & \textbf{DSC(\%) $\downarrow$} & \textbf{PSNR $\uparrow$} & \textbf{SSIM $\uparrow$} \\ 
    \midrule
    EM & 4/255 & 19.88 & 32.55 & 39.04 & 55.97 & 54.20 & 70.02 & 55.58 & 70.69 & 36.72 & 0.9556 \\
    TAP & 4/255 & 21.56 & 34.77 & 36.65 & 53.23 & 36.68 & 53.38 & 54.80 & 70.19 & 36.80 & 0.9528 \\
    LSP & 4/255 & 20.72 & 34.16 & 57.45 & 72.65 & 48.89 & 65.39 & 57.28 & 72.17 & 37.62 & 0.9795 \\
    AR & 4/255 & 25.81 & 40.39 & 54.74 & 70.37 & 53.29 & 69.20 & 56.83 & 72.06 & 41.65 & 0.9899 \\ 
    SEP & 4/255 & 22.05 & 35.47 & 36.50 & 54.37 & 46.41 & 62.98 & 55.44 & 70.76 & 36.47 & 0.9354 \\
    \midrule
    \textbf{UMed} & 4/255 & 0.46 & 0.92 & 32.36 & 48.81 & 39.48 & 56.43 & 56.74 & 71.91 & \textbf{51.65} & \textbf{0.9964} \\
    \textbf{UMed} & 8/255 & 0.04 & 0.08 & 10.56 & 19.10 & 16.82 & 28.49 & 56.10 & 71.09 & 44.63 & 0.9862 \\
    \textbf{UMed} & 12/255 & \textbf{0.03} & \textbf{0.04} & \textbf{9.46} & \textbf{17.21} & \textbf{17.58} & \textbf{29.08} & \textbf{0.19} & \textbf{0.38} & 41.15 & 0.9751 \\
    \bottomrule
    \end{tabular}
}
\caption{Robustness against different defenses on BUSI. We use U-Net as the MIS model. The kernel size of Gaussian blur is $3\!\times\!3$. The quality of JPEG compression is 60. The bound of adversarial training is $\frac{4}{255}$ in terms of $\ell_\infty$-norm.}
\label{tab:robust}

\end{table*}

\begin{table}[t]
    \centering
    \setlength\tabcolsep{5.4pt}
    \scalebox{0.78}{
    \begin{tabular}{@{}ccccccc@{}}
    \toprule
     & \multicolumn{2}{c}{\textbf{BUSI}} & \multicolumn{2}{c}{\textbf{Chest X-ray}} & \multicolumn{2}{c}{\textbf{Kvasir-SEG}}  \\ 
     \cmidrule(lr){2-3}\cmidrule(lr){4-5}\cmidrule(lr){6-7}
    \multirow{-2}{*}{\textbf{Protector}} & \textbf{PSNR $\ \uparrow$} & \textbf{SSIM$\ \uparrow$} & \textbf{PSNR$\ \uparrow$} & \textbf{SSIM$\ \uparrow$} & \textbf{PSNR $\ \uparrow$} & \textbf{SSIM$\ \uparrow$}  \\ \midrule
    EM & 36.72 & 0.9556 & 38.42 & 0.9483 & 38.43 & 0.9375 \\
    TAP & 36.80 & 0.9528 & 38.92 & 0.9527 & 37.50 & 0.9501 \\
    LSP & 37.62 & 0.9795 & 33.39 & 0.9526 & 34.22 & 0.9303 \\
    AR & 41.65 & 0.9899 & 33.53 & 0.9552 & 34.28 & 0.9337 \\
    SEP & 36.47 & 0.9354 & 36.50 & 0.9066 & 36.75 & 0.8766 \\
    \textbf{UMed} & \textbf{51.65} & \textbf{0.9964} & \textbf{46.73} & \textbf{0.9962} & \textbf{51.71} & \textbf{0.9963} \\ \bottomrule
    \end{tabular}
    }
\caption{Quantitative measure of invisibility on three MIS datasets, \ie BUSI, Chest X-ray, and Kvasir-SEG, in terms of PSNR and SSIM between the protected images and original images.}
\label{tab:invis}
\end{table}

\subsection{Analysis on the Invisibility}
{
In Table \ref{tab:invis}, the PSNR and SSIM of UMed outperform existing protection methods. This advantage is attributed to our strategy of concentrating perturbations specifically on the key textures within the ROI and contours of ROI. 
As shown in Fig.~\ref{fig:vis}, our UMed only generates a small amount of perturbations. Meanwhile, the perturbations we inject have a distribution more similar to the intensity of the ROI's contours ($\mathcal G_c$) and textures ($\mathcal G_t$), rendering them less perceptible. 
These strategies not only avoid obvious visual distortions but also ensure that the image's intrinsic characteristics are retained.
}

\subsection{Analysis on the Robustness to Defenses}
UEs are often sensitive to degradation, \textit{e.g.,} JPEG compression~\cite{defence}, and empirical defense strategies, \textit{e.g.,} adversarial training~\cite{at}.
To evaluate the robustness of these methods, we allow the data exploiters' models to adopt various types of distortions. The distortions include Gaussian filtering (kernel size $3\!\times\!3$), JPEG compression (quality 30), and adversarial training (perturbation bound $\ell_\infty=\frac{4}{255}$). 
As shown in Table \ref{tab:robust}, our UMed under the same perturbation bound ($\frac{4}{255}$) as other methods, achieves comparable protective effects with the best method TAP against various distortions. 
Notably, in the pursuit of invisibility, UMed adds much fewer perturbations. 
When we increase the bound to $\frac{8}{255}$ or $\frac{12}{255}$, the robustness of UMed significantly improves, surpassing existing methods. 
Meanwhile, the invisibility metrics, PSNR and SSIM, achieved by our UMed are still outstanding. 
This represents a trade-off between robustness and invisibility: enhancing robustness requires more perturbations to resist corruption, whereas improving invisibility necessitates reducing perturbations, thus increasing the vulnerability to distortion methods. 
In practical applications, we can adjust the invisibility and robustness of UMed based on specific scenarios.

\subsection{Ablation Study}
In Table \ref{tab:abla}, the performance of UMed significantly drops when it lacks either the contour or texture perturbations (Cases B and E). This is attributed to the dependency of MIS models on contour and texture features for predictions. Perturbing only one type of feature leaves the possibility of learning the other useful features for ROI segmentation. 
Meanwhile, compared to the vanilla U-Net (Case C), the CDC U-Net (Case B) achieves better protective performance. This suggests that the CDC can improve the U-Net-based generator's capability in capturing and enhancing the differences between the differences between contour pixels and their neighbors. MIS models, when fitted with these enhanced contour perturbations, cannot effectively learn the original contour information present in the images.
The comparison between Cases E and F shows that the guidance of texture information boosts the protective efficacy of UMed. The adaptive bound (lower than the fixed $\epsilon$) guided by LBP feature maps, facilitates the generation of perturbations that are more easily learnable and possess a distribution more similar to the original texture (refer to Fig.~\ref{fig:vis}).
Additionally, to ensure whether UMed's robust protective performance is solely due to the addition of perturbations to contours or within the ROI, we present the performance of EM with restricted perturbation regions in Cases D and G. The results in Table \ref{tab:abla} demonstrate that this restriction does not effectively improve protection capabilities for the MIS dataset.
This also strongly supports our claim that the smartly designed prior-aware perturbations help in achieving better performance and transferability.

\begin{table}[t]
\centering
\resizebox*{1.0 \linewidth}{!}{
    \begin{tabular}{@{}cccccc@{}}
    \toprule
    \textbf{Case} & \textbf{Protector} & \textbf{\begin{tabular}[c]{@{}c@{}}Contour\\ Perturbator\end{tabular}} & \textbf{\begin{tabular}[c]{@{}c@{}}Texture\\ Perturbator\end{tabular}} & \textbf{Jac.} & \textbf{DSC} \\ 
    \midrule
    A & UMed & $\mathcal G_c$ w/ CDC U-Net & $\mathcal G_t$ w/ LBP & \textbf{3.45} & \textbf{6.34} \\ \midrule
    B & UMed & $\mathcal G_c$ w/ CDC U-Net & - & 9.08 & 15.13 \\
    C & UMed & $\mathcal G_c$ w/ vanilla U-Net & - & 15.17 & 22.73 \\
    D & EM & Contour only & - & 29.47 & 42.43 \\ \midrule
    E & UMed & - & $\mathcal G_t$ w/ LBP & 15.51 & 24.20 \\
    F & UMed & - & $\mathcal G_t$ w/o LBP & 17.66 & 25.85 \\
    G & EM & - & ROI only & 19.91 & 31.96 \\ \bottomrule
    \end{tabular}
}
\caption{Ablation results (\%) on BUSI. We report the average clean Jac. $\downarrow$ and DSC $\downarrow$ of four MIS models, \ie U-Net, Attention U-Net, U-Net++, and TransUNet.}
\label{tab:abla}
\end{table}

\section{Conclusion}
We propose a novel UEs method for protecting MIS datasets, namely UMed, which makes unauthorized MIS models unable to learn useful information from the protected datasets. 
\blue{
UMed first explores the feasibility of utilizing contour and texture features, which are important prior knowledge in MIS, to generate UEs. 
This enables the creation of more easily learnable contour and texture shortcuts, requiring perturbations only within the ROI and its contour rather than across the entire image. 
As a result, UMed achieves superior protective capability and transferability with fewer perturbations, making the protection more invisible and effective. 
Overall, UMed offers an effective method for MIS dataset protection, encouraging more institutions to publicly share their MIS datasets by alleviating concerns about unauthorized model training. 
Moreover, the strategy of UMed to perturb critical features provides novel insights into the generation of UEs.
}

\section*{Acknowledgement}
This research was done at the Rapid-Rich Object Search (ROSE) Lab, Nanyang Technological University, and supported by the National Key Research and Development Program of China (Grant No.2022YFB3207700), the National Natural Science Foundation of China (Grants No.62272022, 62306061, and 62331006), the NTU-PKU Joint Research Institute (sponsored by the Ng Teng Fong Charitable Foundation), the Science and Technology Foundation of Guangzhou Huangpu Development District (Grant No.2022GH15), and the Guangdong Provincial Regional Joint Fund - Regional Cultivation Project (Grant No.2023A1515140037)
\bibliographystyle{named}
\bibliography{ijcai24}

\end{document}